\documentclass[preprint, showpacs,preprintnumbers,amsmath,amssymb,nofootinbib]{revtex4}

 \usepackage{epsf}
 \usepackage{graphicx}
\usepackage{float}
\begin{document}
\newcommand{\beq}[1]{\begin{equation}\label{#1}}
 \newcommand{\eeq}{\end{equation}}
 \newcommand{\bea}{\begin{eqnarray}}
 \newcommand{\eea}{\end{eqnarray}}
 \def\disp{\displaystyle}

 \def\gsim{ \lower .75ex \hbox{$\sim$} \llap{\raise .27ex \hbox{$>$}} }
 \def\lsim{ \lower .75ex \hbox{$\sim$} \llap{\raise .27ex \hbox{$<$}} }

\title{\Large \bf The distance duality relation and the temperature profile of Galaxy Clusters}
\author{Shuo Cao and Zong-Hong Zhu$^\ast$}

\address{ Department of Astronomy, Beijing
Normal University, Beijing, 100875, China }
\email{zhuzh@bnu.edu.cn}

\begin{abstract}
The validity of distance duality relation,
$\eta=D_L(z)(1+z)^{-2}/D_A(z)=1$, an exact result required by the
Etherington reciprocity theorem, where $D_A(z)$ and $D_L(z)$ are the
angular and luminosity distances, plays an essential part in
cosmological observations and model constraints. In this paper, we
investigate some consequences of such a relation by assuming $\eta$
a constant or a function of the redshift. In order to constrain the
parameters concerning $\eta$, we consider two groups of cluster gas
mass fraction data including 52 X-ray luminous galaxy clusters
observed by Chandra in the redshift range $0.3\sim 1.273$ and
temperature range $T_{\rm gas}> 4$ keV, under the assumptions of two
different temperature profiles [1]. We find that the constant
temperature profile is in relatively good agreement with no
violation of the distance duality relation for both
parameterizations of $\eta$, while the one with temperature gradient
(the Vikhlinin et al. temperature profile) seems to be incompatible
even at 99\% CL.

\end{abstract}


 \maketitle
 \renewcommand{\baselinestretch}{1.5}

\section{Introduction}\label{sec:introduction}
%%%%%%%%%%%%%%%%%%%%%%%%%%%%%%%%%%%%%%%%%%%%%%%%%%%%%%%%%%%%%%%%%%%%%%%%%%%%%%
%%%%%%%%%%%%%%%%%%%%%%%%%%%%%%%%%%%%%%%%%%%%%%%%%%%%%%%%%%%%%%%%%%%%%%%%%%%%%%

The reciprocity relation [2,3], $ D_{ L}/D_A(1+z)^{-2}= 1$, also
known as reciprocity law or reciprocity theorem, which links the
luminosity distance $D_L(z)$ with the angular diameter distance
$D_A(z)$ at redshift z, is a fundamental result for astronomical
observations in cosmology. It is extensively used in various domains
ranging from gravitational lensing studies to analyses of the cosmic
microwave black body radiation observations, as well as for galaxy
and galaxy cluster observations [4-22].

As is known to everyone, one fundamental hypothesis in General
Relativity is that light travels along null geodesics in a
Riemannian spacetime (see [23] for instance), based on which the
reciprocity law is generally established. Actually, the reciprocity
relation is also valid for other cosmological models based on
Riemannian geometry besides the FLRW cosmology. The only condition
is that sources and observers are connected by null geodesics in a
general Riemannian spacetime. Up to now, diverse astrophysical
mechanisms such as gravitational lensing and dust extinction have
been proved to be capable of causing obvious deviation from the
distance duality [24] and testing this equality with high accuracy
can also provide a powerful probe of exotic physics [25]. Therefore,
the verification of the observational validity of the reciprocity
law is, perhaps, one of the major challenging open problems in
modern cosmology. Due to this, many previous works have been done
regarding some attempts to test the identity with diverse model
parameterizations of $\eta$ [25-27]. In this paper, we assume that
the reciprocity relation can be a constant or a function of the
redshift, namely:

\begin{equation}
  \frac{D_{\scriptstyle L}}{D_{\scriptstyle A}}{(1+z)}^{-2}= \eta_c , \\
  \frac{D_{\scriptstyle L}}{D_{\scriptstyle A}}{(1+z)}^{-2}= \eta(z),
  \label{rec}
\end{equation}
where $\eta_c$ denotes a constant around the standard case
($\eta_c=1$) and $\eta(z)$ quantifies a possible time-dependent
departing from the standard result ($\eta=1$).

For instance, Bassett \& Kunz (2004) [25] used current supernovae Ia
data as a measurement of the luminosity distance $D_L$, and
estimated $D_A$ from FRIIb radio galaxies [28] and ultra compact
radio sources [9,29,30] to detect possible new physics.

Meanwhile, observations from Sunyaev-Zeldovich effect (SZE) and
X-ray surface brightness from galaxy clusters may also provide us a
test for the distance duality relation. In order to quantify the
$\eta$ parameter, Uzan et al. (2004) [26] fixed $D_A(z)$ by using
the cosmic concordance $\Lambda$CDM model [31] with $D_{A}(z)$ from
18 galaxy clusters [32]. They got $\eta = 0.91^{+ 0.04}_{-0.04}$
(1$\sigma$) by assuming $\eta$ a constant. More recently, De
Bernardis et al. (2006) [33] have also carried out semblable works
using the angular diameter distances from 38 galaxy clusters
provided by the sample of Ref. [34], and they obtained $\eta
=0.97^{+0.03}_{-0.03}$ at $1\sigma$ in the context of $\Lambda$CDM
model. More recently, Holanda et al. (2010) [27] used the
reciprocity relation to show how one may assess the cluster geometry
by using the constraint provided by the reciprocity relation, with
the final result that the elliptical geometry for clusters as
advocated by De Filippis et al. is more favorable.

In this paper, instead of testing the reciprocity relation, we take
it for granted in order to show how we may assess the cluster
temperature profiles by using the constraint provided by the
reciprocity relation. A robust method is that we could obtain the
observational $\eta(z)$ at redshift $z$ from the cluster gas mass
fraction, $f_{\rm gas} = M_{\rm gas}/M_{\rm tot}$, which could be
inferred from X-ray observations of clusters of galaxies. Notice
that the presence of temperature profiles plays an essential part in
the determination of measured total mass, the fraction of mass in
stars and, hence, the gas fraction. In this context, with two groups
of cluster gas mass fraction data from the Chandra satellite [1], we
test the influence on the distance duality relation when different
assumptions about the cluster temperature profiles are considered.
Our analysis here will be based on two parametric representations
defined by Eq.(\ref{rec}), namely: I. $\eta= \eta_c$; II. $\eta (z)
= 1 + \eta_{a} z$. The first expression is on the assumption that
$\eta$ does not evolve with redshifts while the second is a
continuous and smooth linear one-parameter expansion.

Obviously, the above parameterizations are inspired by similar
expressions for $w(z)$-the equation of state in the constant $w$ and
time-varying dark energy models (see Ref.[17,35,36]). More
importantly, for our subsequent analysis, concerning a given data
set, the likelihood of $\eta_c$ or $\eta_a$ must be peaked (or
$\Delta\chi^2$ minimizes equivalently) at $\eta_c=1$ or $\eta_a=0$
in order to satisfy the Etherington theorem. As we shall see,
assuming the strict validity of the reciprocity theorem, our
analysis indicates that the constant temperature assumption tends to
be more compatible with no violation of the reciprocity relation. In
principle, this kind of result is an interesting example of how a
cosmological condition correlates to the local physics.

This paper is organized as follows. In Section~\ref{sec:Sample} we
present two samples of cluster gas mass fraction data from 52 X-ray
luminous galaxy clusters obtained by Chandra. We then briefly
describe the analysis method and show the results of constraining
parameters including $\eta_c$ and $\eta_a$ in
Section~\ref{sec:analysis}. Finally, we summarize our main
conclusions in Section~\ref{sec:Conclusions}.

\section{Galaxy Cluster Samples }\label{sec:Sample}

In order to constrain the possible values of $\eta_{c}$ and
$\eta_{a}$, we consider two samples of cluster gas mass fraction
data from 52 X-ray luminous galaxy clusters obtained by Chandra in
the redshift range $0.3\sim 1.273$ and temperature range $T_{\rm
gas}> 4$ keV, one of which is estimated by assuming a constant
temperature equal to $T_{\rm gas}$ [1], while the other is on the
hypothesis of a different temperature profile which reproduces well
the temperature gradients in nearby relaxed systems [37]:
\begin{equation}
T(r) = 1.23 \frac{ (x/0.045)^{1.9}+0.45 }{(x/0.045)^{1.9}+1}
  \frac{T_{\rm gas}}{\left[ (x/0.6)^2+1 \right]^{0.45}},
\label{eq:tr}
\end{equation}
where $x = r /R_{500}$. Moreover, Ettori et al. (2009) [1] found
this temperature profile might increase $f_{\rm gas}(<R_{500})$ by
about 16 percent with respect to the estimates obtained under the
assumption of isothermality. Because of this, in the following
analysis, we make use of the gas mass fraction data obtained with
different temperature profiles to limit parameters. We recall here
that in the overall sample of 52 objects at $z>0.3$, $(H_0,
\Omega_{\rm m}=1-\Omega_{\Lambda}) = (70 km s^{-1}Mpc^{-1}, 0.3)$ is
assumed, which is also used in the computations throughout this
paper.

\begin{figure}
\begin{center}
\includegraphics[width=0.5\hsize]{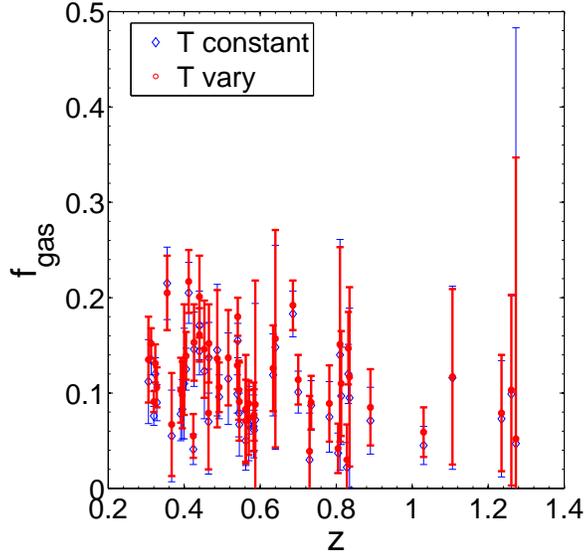}
\end{center}
\caption{ The gas mass fraction data [1]. The open blue diamonds and red circles with the associated error bars stand, respectively, for the constant and Vikhlinin et al. temperature profile samples.
\label{T}}
\end{figure}

In Fig. 1 we plot two samples of cluster gas fraction data from 52
X-ray luminous galaxy clusters obtained in the redshift range
$0.3\sim 1.273$ [1], which were also used as a proxy for the
cosmological parameters such as $\Omega_{\rm m}$ and $w$ [1].
However, since these samples are endowed with different temperature
profile assumptions, our major interest here is to confront these
underlying hypotheses with the validity of the reciprocity relation.

Thereafter, our attention is paid to the process of extracting
$\eta$ from the gas mass fraction. As is well known, old, relaxed,
rich galaxy clusters should provide us with a characteristic sample
of the matter content of the universe if they are large enough.
Moreover, the baryonic fraction $f_{gas}$, should remain the same
during cosmic evolution if the baryonic-to-total mass ratio in
clusters is equal to the cosmological baryonic mass fraction ratio
$\Omega_{\rm b}/\Omega_{\rm m}$. Since clusters are observed at
different redshifts and the reconstructed $f_{gas}$ depends on the
assumed distance to the cluster, this data can be used to constrain
$\eta$. Following Allen et al. (2008) [38], we compute the gas mass
fraction given by

\begin{equation}
\label{eq:fraction}
f_{\rm gas}(z)=\frac{KA\gamma b}{1+s}\frac{\Omega_{\rm b}}{\Omega_{\rm m}}\left(\frac{d_{\rm A}^{\rm ref}(z)}{d_{\rm A}(z)}\right)^{1.5},
\end{equation}
\noindent where five parameters ($K$, $A$, $\gamma$, $b$, $s$) are
related to modeling the cluster gas mass fraction. For example, the
factor $A$ is the angular correction factor and is close to unity
for all redshifts [38].

\begin{table}
\begin{center}
\begin{tabular}{c|c|c|c}\hline\hline
 Cluster & \hspace{4mm} Parameter\hspace{4mm}  & \hspace{4mm}Allowance\hspace{4mm}\\ \hline

 Calibration/modelling & K & $1.0 \pm 0.1$\\
 Non-thermal pressure & $\gamma$ & $1.0 < \gamma <1.1$\\
Gas depletion & b & $0.874 \pm 0.023$\\
 Stellar mass & s & 0.18-0.012$T_{gas}$ \\
 Standard $\Omega_b h^2$ & $\Omega_b h^2$ & $0.0233 \pm 0.0008$ \\
 $\Omega_m$& $\Omega_m$ & 0.3 \\
 Hubble constant & $H_0$ & $70 km s^{-1}Mpc^{-1}$ \\
 \hline\hline
\end{tabular}
\end{center}
\caption{\label{tab1} Summary of the standard systematic allowances
and priors included in the Chandra $f_{gas}$ analysis.}

\end{table}

The depletion parameter $b$ indicates the amount of cosmic baryons
thermalized within the cluster potential. This ``bias'' factor was
modeled as $b=b_0(1+a_{\rm b}z)$ with the uniform priors
$0.65<b_0<1.0$ and $-0.1<a_{\rm b}<0.1$ suggested by cosmological
simulations [38]. More recently, according to various SPH and
Eulerian simulations of a single cluster presented in the Santa
Barbara Comparison Project, Frenk et al. (1999) [39] gained $b =
0.92 \pm 0.07$. We adopt, in our paper, the result from the
simulated dataset in Ref. [1]: $b_{500} = 0.874 \pm 0.023$, which is
only related to the most massive systems, subjected only to the
gravitational heating, high-temperature objects. The parameter
$s=f_{\rm star}/f_{\rm gas}$ denotes the baryon gas mass fraction in
stars. Allen et al. (2008) [38] modeled it as $s=s_0(1+a_{\rm s}z)$,
using the uniform prior $-0.2<a_{\rm s}<0.2$ as well as the Gaussian
prior $s_0=0.13\pm0.01$. However, recent works found that an
additional intracluster light (ICL) at very low surface brightness
might also contribute to the total cold baryonic content of galaxy
clusters: $f_{cold} = f_{star} + f_{ICL}$ [40,41]. More recently,
Lagana et al. (2008) [42] discussed in detail the baryonic content
of five massive galaxy clusters, including an ICL contribution. They
concluded that the stellar-to-gas mass ratio relating to the gas
temperature could be expressed through the relation: $ f_{\rm cold}
= f_{\rm star} + f_{\rm ICL} = \left( 0.18 - 0.012 T_{\rm gas}
\right) f_{\rm gas}$, where $T_{gas}$ is measured in keV and
$T_{gas}>4$ keV is one selection criterion. This parametrization was
extensively discussed in Ref. [1] and is also adopted in our work.
$\Omega_b$ is the present dimensionless density parameter of the
baryonic matter and the WMAP observations give $\Omega_b h^2 =
0.0233 \pm 0.0008$ [31]. $D_{\rm A}^{\rm ref}$ is the angular
diameter distance computed in a reference, spatially-flat
$\Lambda$CDM model with $\Omega_\Lambda=0.7$ ($\Omega_m=0.3$), and
$D_{\rm A}$ is the true angular diameter distance. The complete set
of standard priors and allowances included in the gas mass fraction
($f_{gas}$) analysis are summarized in Table \ref{tab1}.

With $\eta^{2}_{obs}=D^{Cluster}_{A}(z)/D^{Th}_{A}(z)$ (see Ref.
[26]), Eq.(\ref{eq:fraction}) could be rewritten as:

\begin{equation}
\label{eq:eta2}
\eta_{obs}^{3}=\frac{KA\gamma b(z)}{1+s(z)}\frac{\Omega_{\rm b}}{\Omega_{\rm m}}f_{\rm gas}^{-1} .
\end{equation}
In our statistical analysis (see the next section), the
observational $\eta^3_{obs}$ and its corresponding uncertainty are
also calculated through Eq.(\ref{eq:eta2}).

\section{Analysis and Results}\label{sec:analysis}

Let us now estimate the $\eta_{c}$ and $\eta_{a}$ parameters for
both samples in two parameterizations for $\eta
=D_{L}(z)(1+z)^{-2}/D_{A}(z)$, e.g., $\eta = \eta_{c}$, $\eta(z) =
1+\eta_{a}z$. To begin with,  we evaluate the likelihood
distribution function $e^{-\chi^{2}/2}$ , where
\begin{equation}
\label{chi2} \chi^{2} = \sum \frac{{\left[\eta^{3} -
\eta^{3}_{obs} \right] }^{2}}{\sigma^2_{\eta^{3}_{obs}} },
\end{equation}
with $\eta^{3}_{obs}$ gained from the $f_{gas}$ data
[Eq.(\ref{eq:eta2})]. As for $\sigma_{\eta^{3}_{obs}}$, the
uncertainty of $\eta^{3}_{obs}$, is calculated through the
propagation of uncertainty statics with a combination of the
observational $f_{gas}$, the best fit values of parameters such as
$K$, $\gamma$ as well as their corresponding standard systematic
allowances (Fig. 1 and Table \ref{tab1}).

\begin{figure}
\begin{center}
\includegraphics[width=0.5\hsize]{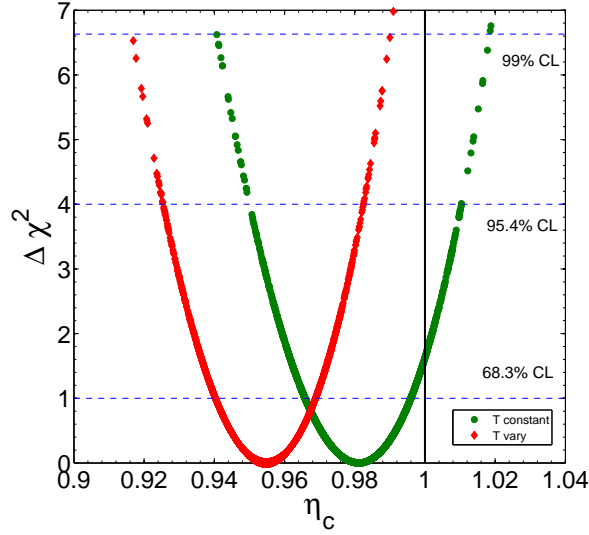}
\end{center}
\caption{The $\eta_{c}-\Delta\chi^2$ plane for the parameterization $\eta=\eta_c$. The filled (green) circles and filled (red) diamonds stand for the constant and Vikhlinin et al. temperature profile samples, respectively.
\label{1}}
\end{figure}

\begin{figure}
\begin{center}
\includegraphics[width=0.5\hsize]{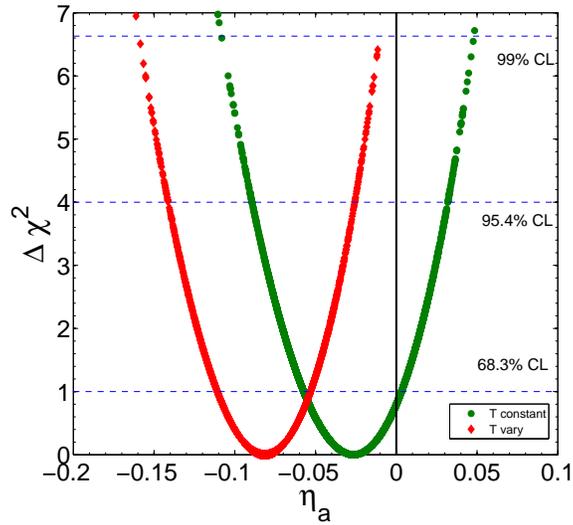}
\end{center}
\caption{The $\eta_{a}-\Delta\chi^2$ plane for the parameterization $\eta(z)=1+\eta_az$. The denotation of different data is the same as Fig.\ref{1}.
\label{2}}
\end{figure}

\begin{figure}
\begin{center}
\includegraphics[width=0.5\hsize]{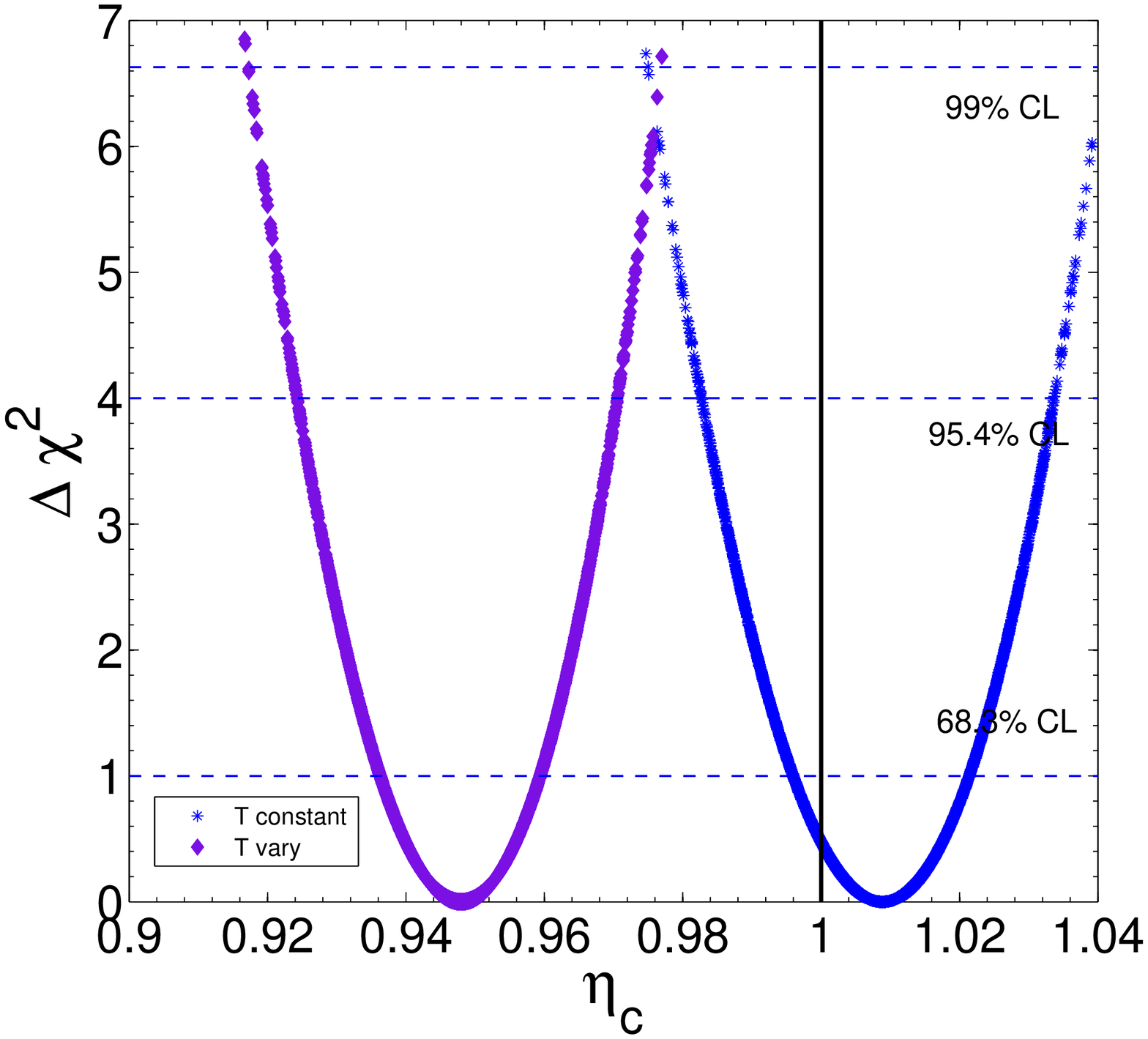}
\includegraphics[width=0.5\hsize]{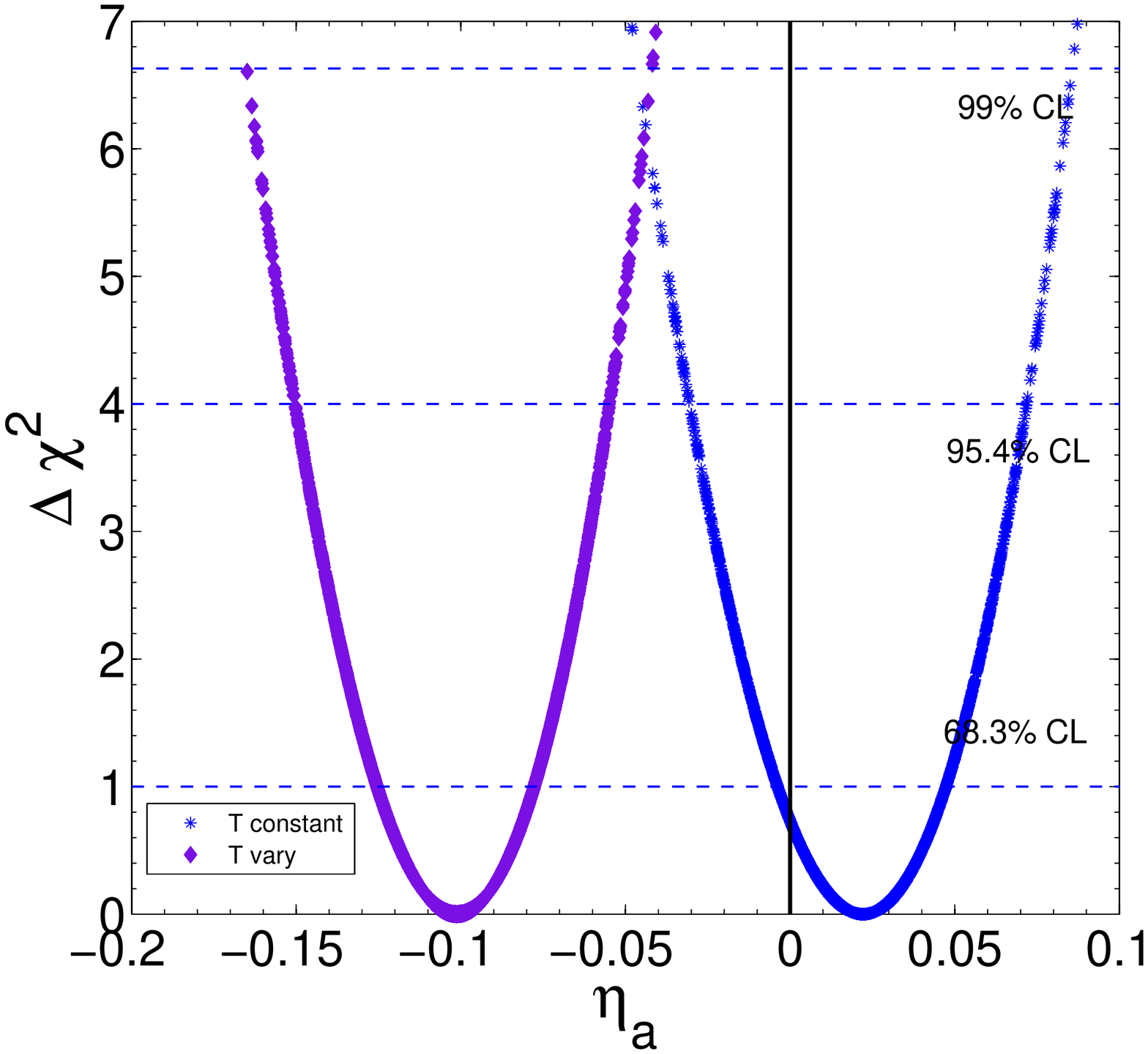}
\end{center}
\caption{The $\eta_{c}-\Delta\chi^2$ planes for two parameterizations $\eta=\eta_c$ and $\eta(z)=1+\eta_az$ with marginalized $f_{gas}$ parameters.
The blue stars and filled purple diamonds stand for the constant and Vikhlinin et al. temperature profile samples, respectively.
\label{2.1}}
\end{figure}

%\begin{figure}
%\begin{center}
%\includegraphics[width=0.7\hsize]{revised2.eps}
%\end{center}
%\caption{The $\eta_{a}-\Delta\chi^2$ plane for the parameterization $\eta(z)=1+\eta_a(z)z$ with marginalized $f_{gas}$ parameters. The denotation of different data is the same as Fig.\ref{2.1}.
%\label{2.2}}
%\end{figure}

In Fig. 1 and 3, we plot the likelihood distributions on the
$\eta_{c}-\Delta\chi^2$ and $\eta_{a}-\Delta\chi^2$ planes for both
samples. We obtain $\eta_{c} = 0.955^{+ 0.015}_{-0.015}$
($\chi_{d.o.f.}^2 = 47.25/52$) and $\eta_{a} = -0.082^{+
0.029}_{-0.029}$ ($\chi_{d.o.f.}^2 = 49.50/52$) at $68.3\%$ CL for
the Vikhlinin et al. temperature profile sample and $\eta_{c} =
0.981^{+ 0.015}_{-0.015}$ ($\chi_{d.o.f.}^2 = 63.44/52$) and
$\eta_{a} = -0.026^{+ 0.034}_{-0.034}$ ($\chi_{d.o.f.}^2 =
64.28/52$) at $68.3\%$ CL for the constant temperature sample. We
can see that the $\eta$ from the latter is in agreement with no
violation of the reciprocity relation while the former, where a
Vikhlinin et al. temperature profile is assumed to describe the
clusters, is incompatible even at 99\% CL. Therefore, we have not
found any evidence for distance duality violation when the constant
temperature sample is considered. However, the same kind of analysis
shown in Fig. 2 and 3 points to be in contradiction with the
Vikhlinin et al. temperature profile hypothesis assumed in the
Vikhlinin et al. sample. Furthermore, considering that many
parameters listed in Table \ref{tab1} are fixed to their best-fit
values, we perform a MCMC analysis and marginalize over these
nuisance parameters [43]. The constraint results shown in Fig. 4
further confirm our conclusions.

\section{Conclusions}\label{sec:Conclusions}

In this paper, we have explored the consequences of the distance
duality relation $\eta = D_{L}(1+z)^{-2}/D_{A}$ based on two samples
of cluster gas mass fraction data from 52 X-ray luminous galaxy
clusters obtained by Chandra in the redshift range $0.3\sim 1.273$
and temperature range $T_{\rm gas}> 4$ keV [1]. We discussed the
consistency between the strict validity of the distance duality
relation and the assumptions about temperature profiles (the
constant and Vikhlinin et al. temperature profiles) used to describe
the galaxy clusters. The $\eta$ parameter was parametrized in two
different functional forms, $\eta = \eta_{c}$ and $\eta =
1+\eta_{a}z$.

By comparing the constant and Vikhlinin et al. temperature profile
samples, we show that the constant temperature profile is more
consistent with no violation of the distance duality relation in the
context of two groups of cluster gas mass fraction data. In the case
of constant temperature sample (see Fig. 2 and 3) we find $\eta_{c}
= 0.981^{+ 0.015}_{-0.015}$ and $\eta_{a} = -0.026^{+
0.034}_{-0.034}$ for constant and linear parametrizations,
respectively. On the other hand, the Vikhlinin et al. temperature
profile model (see Fig. 2 and 3) seems to be marginally incompatible
with $\eta_{c} = 0.955^{+ 0.015}_{-0.015}$ and $\eta_{a} = -0.082^{+
0.029}_{-0.029}$ for constant and linear parameterizations,
respectively. Our analysis indicates that the constant temperature
assumption tends to be more compatible with the Etherington theorem
at 68.3\% CL (between 68.3\% and 95.4\% CL for the first
parametrization) compared with the Vikhlinin et al. temperature
profile hypothesis, while the latter is incompatible even at 99\%
CL. The results with marginalized $f_{gas}$ parameters (Fig. 4)
further confirm these conclusions.

In summary, we find that, according to the statistical analysis
presented here, the constant temperature profile of clusters tends
to be more consistent with no violation of the distance duality
relation regarding two groups of cluster gas mass fraction data.
This reinforces the interest in the observational search for such
kind of data from clusters at high redshifts. With better data, the
method proposed here based on the validity of the distance duality
relation should lead to improved limits on the temperature profiles
of gas in galaxy clusters.

\section*{Acknowledgments}
This work was supported by the National Natural Science Foundation
of China under the Distinguished Young Scholar Grant 10825313 and
Grant 11073005, and by the Ministry of Science and Technology
national basic science Program (Project 973) under Grant No.
2007CB815401.
%\clearpage

\end{document}